\documentclass[10pt, conference]{IEEEtran}
\usepackage{newcent}
\usepackage{graphicx, color}
\usepackage{amsmath, amssymb}
\usepackage{subfig}
\usepackage{soul}
\usepackage{url}
\usepackage{booktabs}
\usepackage{multirow}
\usepackage{cite}

\begin{document}

\title{\textsc{Divide and Conquer:} \\ Partitioning OSPF networks with SDN}
\author{\IEEEauthorblockN{Marcel Caria, Tamal Das, and Admela Jukan}
\IEEEauthorblockA{Technische Universit\"at Carolo-Wilhelmina zu
Braunschweig\\
Email: \{m.caria, t.das, a.jukan\} @tu-bs.de}
\and
\IEEEauthorblockN{Marco Hoffmann}
\IEEEauthorblockA{NOKIA\\
marco.hoffmann@nsn.com}
}
\maketitle

\begin{abstract}
Software Defined Networking (SDN) is an emerging network control paradigm focused on logical centralization and programmability. At the same time, distributed routing protocols, most notably OSPF and IS-IS, are still prevalent in IP networks, as they provide shortest path routing, fast topological convergence after network failures, and, perhaps most importantly, the confidence based on decades of reliable operation. Therefore, a hybrid SDN/OSPF operation remains a desirable proposition. In this paper, we propose a new method of hybrid SDN/OSPF operation. Our method  is different from other hybrid approaches, as it uses SDN nodes to \emph{partition} an OSPF domain into \emph{sub-domains} thereby achieving the traffic engineering capabilities comparable to full SDN operation. We place SDN-enabled routers as sub-domain border nodes, while the operation of the OSPF protocol continues unaffected. In this way, the SDN controller can tune routing protocol updates for traffic engineering purposes before they are flooded into sub-domains. While local routing inside sub-domains remains stable at all times, inter-sub-domain routes can be optimized by determining the routes in each traversed sub-domain. As the majority of traffic in non-trivial topologies has to traverse multiple sub-domains, our simulation results confirm that a few SDN nodes allow traffic engineering up to a degree that renders full SDN deployment unnecessary.
\end{abstract}

\section{Introduction}\label{Introduction-section}
Distributed IP routing protocols, like OSPF or IS-IS, have worked consistently and predictably in the current Internet and have proven their reliable operation over time. Software Defined Networking (SDN), on the other hand, is a new networking paradigm based on a logically centralized and programmable control plane, that has gained a lot of attention recently. In fact, most network equipment vendors have announced the intention to build devices that support the OpenFlow protocol, the de facto SDN messaging standard, or have already released OpenFlow-capable products~\cite{onfMemberlist}.

Migrating to a fully SDN-enabled operation is however not without issues and new costly investments. In fact, ISPs are still reluctant regarding the change of the control plane paradigm in their networks from \emph{distributed} to \emph{centralized}, as distributed routing protocols operate consistently and predictably over years, efficiently control real life conditions, and reliably recover from network failures. A migration to SDN, on the other hand, requires new hardware, new tools, and new expertise for network administrators, while SDN still fights with some hard-to-kill scalability preconceptions~\cite{scalability}. It is therefore no surprise that a lot of work has recognized the importance of the so-called SDN/OSPF hybrid networking paradigm~\cite{hybrid_1, hybrid_2, hybrid_3, hybrid_4} and many OpenFlow devices support a hybrid mode. In the current approaches, the SDN nodes build their regular forwarding tables from OSPF, while the SDN controller can insert higher priority rules (also with more sophisticated matching parameters).

\begin{figure}[tb] \center
\includegraphics[width=8.7cm]{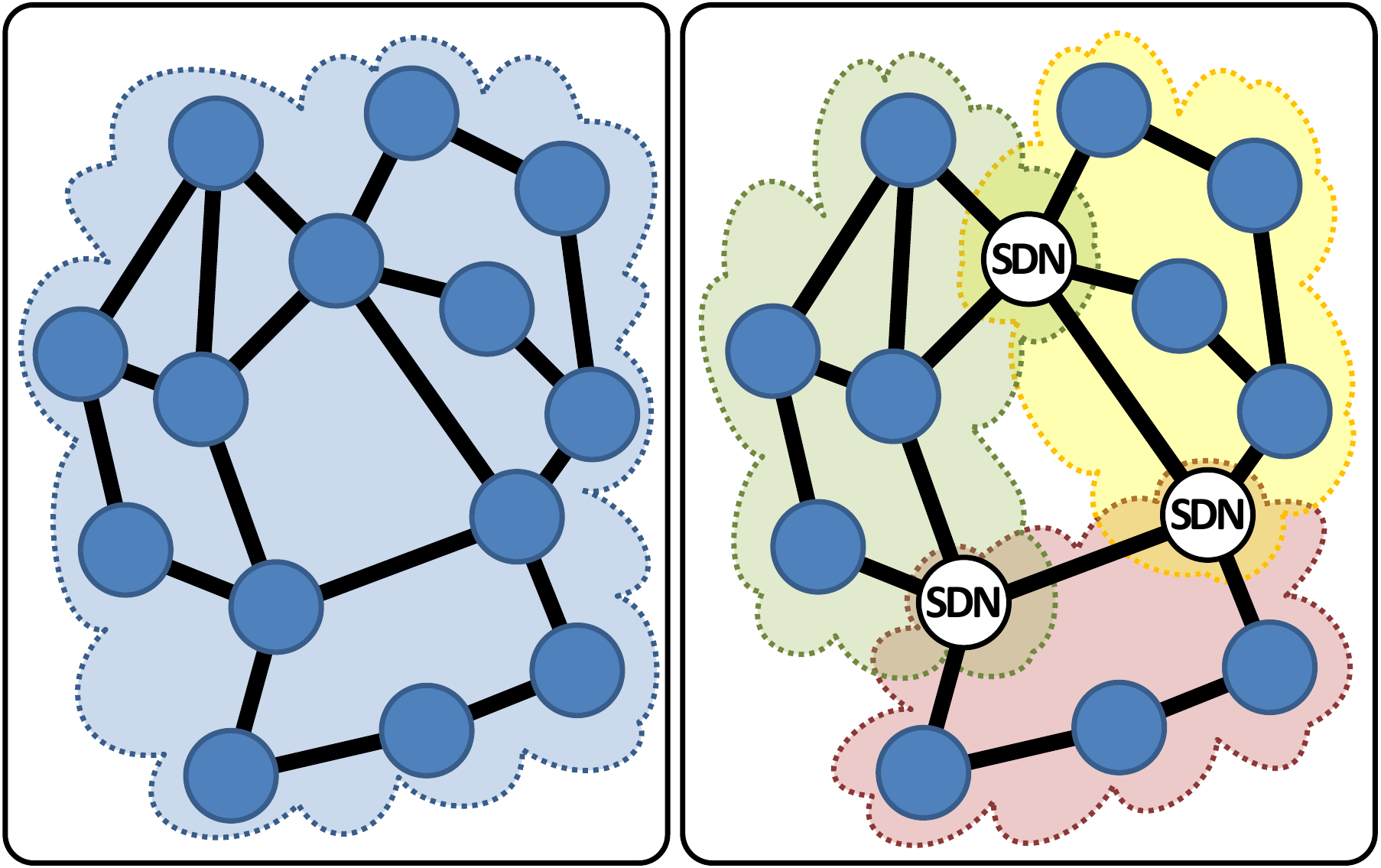}
\caption{Partitioning OSPF with SDN: LSA flooding is limited by SDN nodes, and each SDN participates in the OSPF protocol of their corresponding sub-domains.}
\label{partitioning} \end{figure}

In this paper, we propose a new method for SDN/OSPF hybrid networking. Our  method is fundamentally different from any previous work as we \emph{partition} the initial OSPF domain with SDN nodes into sub-domains, such that routing protocol updates for inter-sub-domain paths can be overridden at SDN border nodes by the SDN controller, e.g., for traffic engineering purposes. This requires that neighboring sub-domains are connected \emph{only} via SDN nodes and do not have any direct links otherwise. The advantage of our idea over other hybrid approaches is that our partitioning method can be implemented into an operational OSPF network by operating the SDN nodes initially (i.e., during the migration) in plain OSPF mode. It follows that the SDN nodes in our scheme belong to \emph{all} sub-domains to which they are connected.

Figure~\ref{partitioning} illustrates the idea. When all sub-domain border nodes are replaced with SDN-enabled devices and an optimized routing has been determined, the SDN nodes start the update process in the individual sub-domains by flooding routing updates that are individually tuned per sub-domain. Our method capitalizes on the advantages of distributed routing protocols by letting the routing \emph{inside} each sub-domain be based on OSPF solely so that it remains unchanged and stable at all times. At the same time, inter-sub-domain routes can be optimized by determining the routes in each traversed sub-domain. As the majority of traffic in non-trivial topologies has to traverse multiple sub-domains, our simulation results confirm that a few SDN nodes allow traffic engineering up to a degree that renders the full SDN deployment unnecessary.

The rest of the paper is organized as follows: Section~\ref{rel-work-section} discusses the related work and Section~\ref{arch-section} presents the network architecture. The mathematical model of the traffic engineering logic is presented in Section~\ref{math-section}. Section~\ref{performance-section} presents the performance study and Section~\ref{conclusion-section} the conclusions.

\section {Related Work}\label{rel-work-section}
\par Our method appears similar to the partitioning of an OSPF domain into so-called OSPF areas~\cite{ospf}, however the use is different, as OSPF areas are used to simplify the administration of large topologies and to reduce the amount of protocol traffic. Our intention is also different, as we partition the OSPF domain to allow SDN-based traffic engineering by controlling the routes of inter-sub-domain traffic.

The SDN/OSPF hybrid networking paradigm has been previously analyzed and explained to a great deal in ~\cite{hybrid_1, hybrid_2, hybrid_3, hybrid_4}. For the capability of the SDN controller to insert higher priority rules into the forwarding tables, this has also been coined as "policy based routing on steroids"~\cite{steroids}. The "topology-based hybrid SDN model" by Vissicchio et al. in~\cite{vissicchio} differs from the other approaches in this respect, and is interesting to us, as it partitions the network into \emph{zones}, whereby each node belongs to one zone only. However, the zone approach is different from our concept of sub-domains, as a zone is defined as a set of interconnected nodes controlled by the same paradigm (i.e., SDN \emph{or} OSPF). In contrary, a sub-domain in our approach is a subgraph of OSPF routers, while the SDN nodes participate in the OSPF.

There are other architectural solutions for traffic engineering (TE), most prominently MPLS. However, we argue that unlike our method, MPLS requires a full deployment in the entire topology and it adds another architectural layer to the network (including its overhead in the data plane). Only a full migration to SDN would actually provide equal performance in terms of TE. Both solutions however, i.e., MPLS and SDN, assume the previously mentioned change to the \emph{centralized} control plane paradigm. In contrast to these solutions, our method relies mainly on OSPF, while optimized routes are \emph{injected} into the regular protocol operation by individually tuned routing protocol updates. Our results also show that our method requires relatively few SDN nodes to achieve the same traffic engineering capability to the one with the full MPLS or SDN deployment.

Finally, an intuitively similar approach would be to partition the initial OSPF domain and connect the sub-domains with BGP, while BGP can then be used for load balancing, like shown in~\cite{bgp}. The degree of freedom here regarding routing optimizations are in fact similar to what our method can offer. However, our method does not require the partitioning of the initial routing domain into multiple \emph{autonomous systems} and provides a clean separation from the BGP setup that operative WANs have already in place.

\section {Network Management Architecture}\label{arch-section}
We propose an SDN/OSPF hybrid network management subsystem, with a TE scheme applicable to a single OSPF routing domain. A routing update is referred to as Link State Advertisement (LSA) and an OSPF router participating in the protocol distributes all its topological information by flooding LSAs throughout the entire network. Figure~\ref{architecturefigure} depicts the assumed Layer~3 network management architecture. As it can be seen, the actual network includes both OSPF and SDN routers. The SDN nodes are connected to the SDN controller and the proposed network management subsystem, referred to as \emph{Hybrid Network Manager} (HNM), uses the controller's northbound API. As our partitioning-based method requires some sort of all-knowing centralized control intelligence, we consider the implementation as a sort of network application on top of the SDN controller. Because the OSPF protocol advertisements are essential here, the used SDN controller must be configurable to simply forward (without any processing) all LSAs received by the SDN nodes to the HNM, as well as to simply forward the LSAs generated by the HNM to the SDN nodes.

\subsubsection{Hybrid Network Manager} 
\begin{figure}[t] \center
\includegraphics[width=8.7cm]{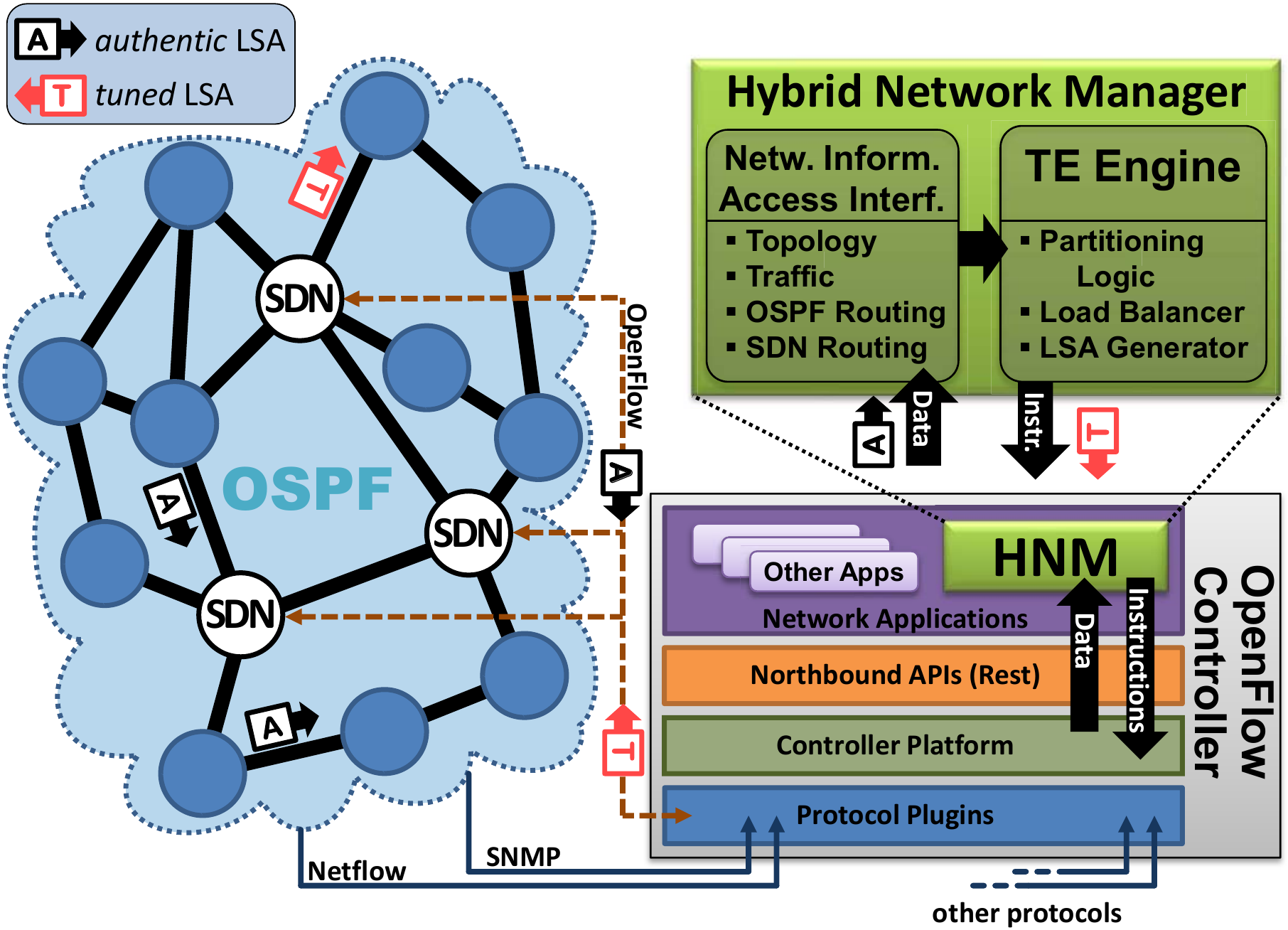}
\caption{The network architecture proposed}
\label{architecturefigure} \end{figure}

To gather all the data relevant to Traffic Engineering (i.e., topology, SDN node placement, traffic, and routing), we assume that the HNM implements a \emph{Network Information Access Interface} that has comprehensive access to various functions and data in the OpenFlow controller, and preprocesses and does the re-formating of this data for the TE Engine, as well as it extracts the routing information from the received LSAs. The \emph{Traffic Engineering Engine} (depicted as "TE Engine" block in Figure~\ref{architecturefigure}) is our main module inside the HNM, which is used to determine the optimum routing based on the provided data. This module is aware of the OSPF topology partitioning, optimizes the inter-sub-domain routing for load balancing, and computes the according (tuned) OSPF link metrics that will be flooded as LSAs into the individual sub-domains. The TE Engine then passes on the computed routes to the SDN controller's \emph{Controller Platform} (shown inside the OpenFlow controller) and forwards the LSAs to the corresponding SDN routers. 

\subsubsection{SDN Integration into an OSPF network} Let us assume a single domain and single area OSPF network. Assuming that a few OSPF routers have been replaced by the SDN-enabled devices, the first step for integration into an OSPF networks is to configure the SDN nodes to operate autonomously in the OSPF mode. When the proposed control and management architecture is fully deployed (i.e., the HNM is operating on top of the SDN controller), the sub-domain border nodes can be switched over to SDN mode. From this time on, the SDN nodes send the \emph{authentic} LSAs (depicted as the black packets marked with an  \emph{A}) received from their OSPF neighbors to the HNM, where they are processed. During the following initialization phase, the HNM is not altering any routing and replies with LSAs according to regular OSPF operation. This phase is used by the HNM to gather all required topology and routing information from the SDN controller and the LSAs received from the OSPF routers. This also involves the logical partitioning of the single OSPF domain into sub-domains depending on the SDN node placement. When the initialization phase is completed (i.e., when the HNM has all data available), the HNM switches over to the load balancing (or, TE) phase, in which it computes the optimum routes that are then translated into link metrics for the \emph{tuned} LSAs (depicted as the red packets marked with a \emph{T}). The HNM computes LSAs for load balancing individually depending on the sub-domain and the advertising border node. It then forwards these LSAs via the OpenFlow channel to the corresponding SDN nodes, which distribute them in the corresponding sub-domain via flooding.

\subsubsection{SDN Node Placement} To enable the separation of the OSPF domain into distinct sub-domains, a few SDN-enabled routers must be placed in strategic positions in the network, such that their removal partitions the topology into disconnected components. Obviously, the way the network is clustered into sub-domains is determined solely by the operator's choice of nodes that will be exchanged with SDN nodes, i.e., the sub-domains are determined only once in the beginning and can be changed only by adding new or changing the position of existing SDN nodes. In graph theory parlance, this means that the SDN-enabled routers must constitute one or multiple vertex separators. An example of how to partition a topology is shown in Figure~\ref{partitioning}, where it can be seen that the removal of any two SDN nodes separates the topology into two disconnected subgraphs. Evidently, the SDN-enabled nodes have to be placed at the \emph{weak points} of the topology from a network reliability perspective. However, it is one of the strengths of our hybrid scheme that in case of a failing sub-domain border node, OSPF takes care of finding alternative paths via other border nodes. We control the routing of inter-sub-domain flows only by determining which border node is \emph{preferred} over the others (by the advertised link metrics), and in case that preferred node fails, the next least cost route via a different border node is automatically used.

\par It should be noted that the partitioning we use in this paper is rather simplistic: For a given number of SDN-enabled routers, we chose those nodes to be replaced that result in the partitioning with the maximum number of inter-sub-domain flows in the network. Better partitioning strategies are in fact likely to yield better results (outside the scope here).

\subsubsection{LSA Tuning} 
\begin{figure}[t] \center
\includegraphics[width=8.7cm]{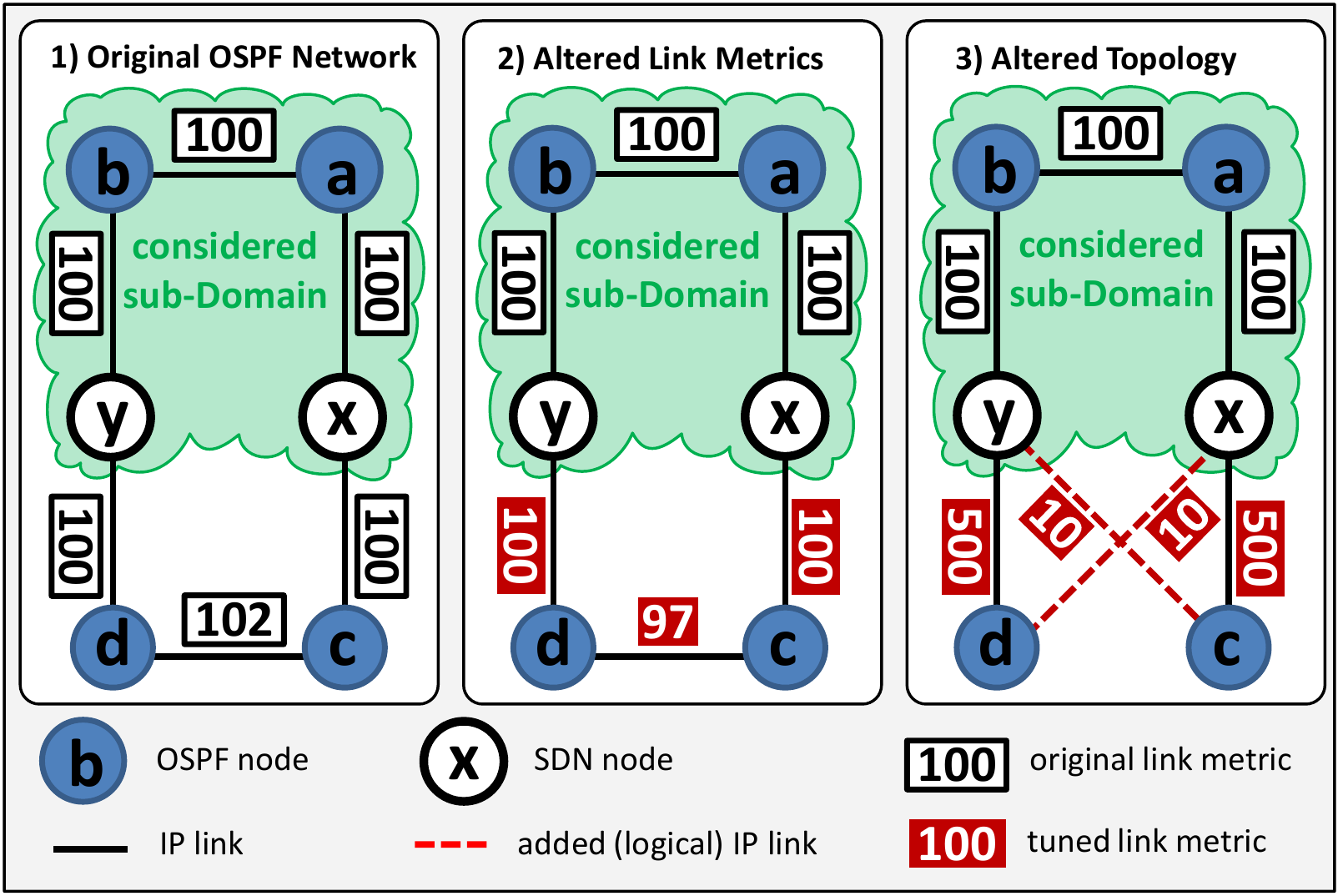}
\caption{Altering the advertised routing information}
\label{lsa} \end{figure}

The partitioning of the OSPF area allows to advertise routing updates customized per sub-domain, which in turn enables to control the routing of traffic flows between the sub-domains. It should be noted that the OSPF routing of sub-domain-internal traffic is not affected at all. Figure~\ref{lsa} depicts how LSA altering can be used to change the routing: The first sub-figure shows the original OSPF network and how the two SDN-enabled nodes $x$ and $y$ are used to partition the domain into two distinct sub-domains (containing the nodes $\{a,b,x,y\}$ in the here considered sub-domain, in which we observe the routing changes, and nodes $\{c,d,x,y\}$ in the other sub-domain respectively). The second sub-figure shows how paths can be controlled, when only the sub-domain-external link metrics are altered (e.g., by changing them to the red numbers) before LSAs are flooded into the considered sub-domain. The third sub-figure shows the more powerful control method, where even the advertised topology is altered (i.e., new \emph{logical} links can be advertised or existing links can be hidden). The corresponding routing paths for inter-sub-domain flows are shown in Table~\ref{flowrouting}, grouped in blocks according to the used LSA modification.

\par It can be seen in Table~\ref{flowrouting} that by altering the link metrics in the LSAs before being flooded to nodes $a$ and $b$ according to the way depicted in second sub-figure of Figure~\ref{lsa}, flows $a\!\rightarrow\! d$ and $b\!\rightarrow\! c$ change their routes so that the link between $a$ and $b$ is relieved and the load on the link between $c$ and $d$ increases instead. This poses just a slight modification, which might come in handy when link $a-b$ is overloaded while link $c-d$ provides enough spare capacity. The third sub-figure of Figure~\ref{lsa} shows the possible routing changes, when even topology information is altered in the flooded LSAs: by advertising the non-existing links $y-c$ and $x-d$, both with low link metrics, all traffic for $c$ exits the upper sub-domain via node $y$ and all traffic for $d$ exits via $x$ (like shown in the third block in Table~\ref{flowrouting}), which wouldn't be possible only by altering link metrics of existing links.

\begin{table}[t]\begin{center}
\begin{tabular}{ c c c }
\toprule
\textbf{LSAs} & \textbf{Traffic Flow} & \textbf{Routing Path} \\
\midrule
\multirow{4}{*}{Original} &
$a\!\rightarrow\! c$	&	$a-x-c$ \\
& $a\!\rightarrow\! d$	&	$a-b-y-d$ \\
& $b\!\rightarrow\! c$	&	$b-a-x-c$ \\
& $b\!\rightarrow\! d$	&	$b-y-d$ \\
\midrule
& $a\!\rightarrow\! c$	&	$a-x-c$ \\
Altered & $a\!\rightarrow\! d$	&	$a-x-c-d$ \\
Link Metrics & $b\!\rightarrow\! c$	&	$b-y-d-c$ \\
& $b\!\rightarrow\! d$	&	$b-y-d$ \\
\midrule
& $a\!\rightarrow\! c$	&	$a-b-y-d-c$ \\
Altered & $a\!\rightarrow\! d$	&	$a-x-c-d$ \\
Topology & $b\!\rightarrow\! c$	&	$b-y-d-c$ \\
& $b\!\rightarrow\! d$	&	$b-a-x-c-d$ \\
\bottomrule
\end{tabular}
\caption{Routing of inter-sub-domain flows in Figure~\ref{lsa} depending on the advertised routing information}\label{flowrouting}
\end{center}\end{table}

\begin{figure*}[t] \center
\includegraphics[width=\textwidth]{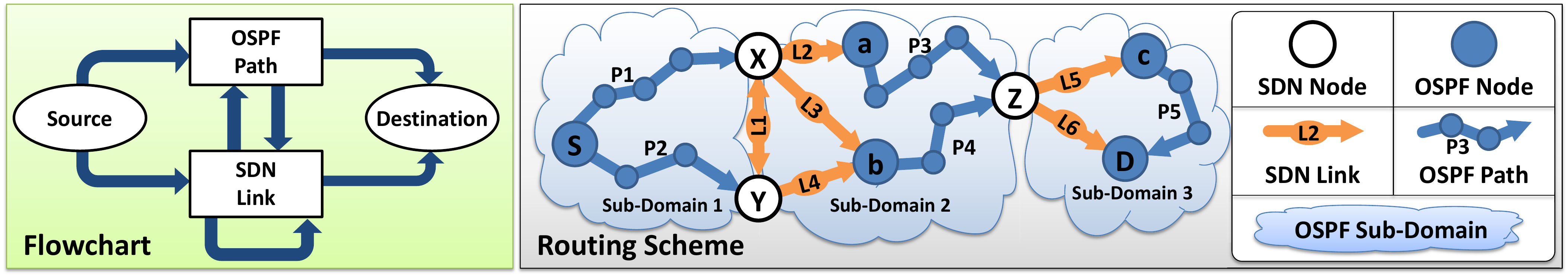}
\caption{Constraints on routing}
\label{routing} \end{figure*}

\section{Load Balancer Model for TE}\label{math-section}
We now present the Integer Linear Programming (ILP) model used for the load balancer in our TE engine. All parameters and variables used are listed in Table~\ref{symbols}. We assume that all OSPF paths inside sub-domains (including their accumulated link metrics) are constant and known. It should be noted that the novelty of the routing scheme in our model requires newly defined constraints on routing which makes the model rather unique and challenging: While OSPF paths must be alternated with SDN links, SDN links alone can be concatenated arbitrarily, like shown in the flowchart of Figure~\ref{routing}.

\vspace{3mm}\noindent\textbf{The routing constraints: }
A valid routing path according to our method is a combination of OSPF paths and SDN links, see the flowchart of Figure~\ref{routing}. Note that our notion of the terms \emph{OSPF path} and \emph{SDN link} differs from the intuitive usage. We define an SDN link as a directional link between two nodes, whereas the source of the link has to be an SDN node and the destination of the link can be any (SDN or OSPF) neighboring node. An OSPF path, on the other hand,  is defined as the \emph{unique least cost path} (whereby the mechanisms like Equal Cost Multi Path are not considered) between a (non-SDN) OSPF node and any other (SDN or OSPF) node \emph{within the same sub-domain}, which also does not traverse any \emph{SDN link} (like defined previously). To illustrate this definition: the directional link from an SDN node to any of its neighbors can be used as an SDN link, while the same can not be part of any OSPF path. 

\begin{table}[t]\begin{center}
\begin{tabular}{ c l }
\toprule
\textbf{Parameter} & \multicolumn{1}{c}{\textbf{Meaning}} \\
\midrule
$\text{N}$ & set of all nodes \\\addlinespace[1.0mm]
$\text{L}\supseteq\text{L}^\text{sdn}$ & set of all links / all SDN links \\\addlinespace[1.0mm]
$\text{SD}$ & set of all OSPF sub-domains\\\addlinespace[1.0mm]
$\text{P}$ & set of all OSPF paths (like defined) \\\addlinespace[1.0mm]
\multirow{2}{*}{$\text{F}\supseteq\tilde{\text{F}}$} & set of all traffic flows / all traffic \\
 & $\text{ }$ flows crossing sub-domain borders \\\addlinespace[1.0mm]
$\text{Y}$ & set of linear cost functions \\\addlinespace[1.0mm]
$\lambda_f$ & traffic demand of flow $f$ \\\addlinespace[1.0mm]
\multirow{3}{*}{$src(x,n)$} & binary: 1 if node $n$ is the \emph{source} of\\
 & $\text{ }$ entity $x$ (which can be a link, a path, \\
 & $\text{ }$ or a flow), and 0 otherwise \\\addlinespace[1.0mm]
\multirow{2}{*}{$dst(x,n)$} & binary: 1 if node $n$ is the \emph{destination} of \\
 & $\text{ }$ entity $x$, and 0 otherwise \\\addlinespace[1.0mm]
\multirow{2}{*}{$tr(p,\ell)$} & binary: 1 if OSPF path $p$ traverses \\
 & $\text{ }$ link $\ell$, and 0 otherwise \\\addlinespace[1.0mm]
$C_\ell$ & capacity of link $\ell$ \\\addlinespace[1.0mm]
\multirow{2}{*}{$u_\ell$} & utilization of link $\ell$ caused by all\\
 & $\text{ }$ intra-sub-domain flows traversing $\ell$ \\\addlinespace[1.0mm]
\multirow{3}{*}{$\ell sa(k,p)$} & binary: 1 if the advertisement of link\\
 & $\text{ }$ metrics according to LSA set $k$ allows \\
 & $\text{ }$  usage of path $p$, and 0 otherwise \\\addlinespace[1.0mm]
\multirow{2}{*}{$bel(k,\alpha)$} & binary: 1 if the LSA set $k$ belongs to \\
 & $\text{ }$ sub-domain $\alpha$, and 0 otherwise \\\addlinespace[1.0mm]
\midrule
\textbf{Variable} & \multicolumn{1}{c}{\textbf{Meaning}} \\
\midrule
\multirow{2}{*}{$R_x(f)$} & binary: 1 if  OSPF path or SDN link $x$ \\
 & $\text{ }$ is used for flow $f$, and 0 otherwise \\\addlinespace[1.0mm]
\multirow{2}{*}{$LSA_k(d)$} & binary: 1 if  LSA set $k$ is advertised \\
 & $\text{ }$ for destination $d$, and 0 otherwise \\\addlinespace[1.0mm]
$U_\ell$ & utilization of link $\ell$\\\addlinespace[1.0mm]
$Cost_\ell$ & utilization cost of link $\ell$\\\addlinespace[1.0mm]
\bottomrule
\end{tabular}
\caption{Summary of Notation}\label{symbols}
\end{center}\end{table}

\par Figure~\ref{routing} shows an example network with the routing scheme proposed. The source node $S$ has exactly one least cost path to each SDN border node ($X$ and $Y$) in Sub-Domain~1. Thus, the route from $S$ to $D$ starts either with OSPF path $P1$ or $P2$, depending on the cost metric for the routing to $D$ advertised by $X$ and $Y$. The next element in the route to $D$ is an SDN link. As an SDN node's flow table can be arbitrarily configured, e.g., for packets matching source addresses of traffic from $S$ and destination addresses of traffic to $D$, we see that $P1$ can be continued with SDN links $L1$, $L2$ or $L3$. In fact,  $P1$ can be continued even with SDN links $L1$ and $L4$ successively, whereby we'd like to note the self loop of the \emph{SDN link} box in the flowchart: Because forwarding in SDN nodes is arbitrarily configurable and not constraint by OSPF, SDN links can be concatenated arbitrarily. Arriving at any of the two first OSPF nodes (node $a$ or node $b$) in Sub-Domain~2, we have no choice for the next element to add to our route to the destination, because each of the two nodes has only a single OSPF path to border node $Z$. Finally, $Z$ can then be configured to forward the packets directly via link $L6$ to $D$, or in case of congestion on that link, forward the packets via $L5$ to OSPF node $c$, from where the packets have to take the OSPF path to the destination. 

\par To assure in our ILP model that there is a valid routing path for each flow, we first require that exactly one forwarding entity (i.e., an OSPF path or an SDN link), which connects to the source node of the said flow, is used.  In other words, routing has to \textbf{start at the source} of the flow:
\[
\forall f\in\tilde{\text{F}}, \,\,\, \forall n\in\text{N} \,\, \text{ with } \,\, src(f,n)=1:
\]\[
\sum_{p\in\text{P}} src(p,n) \cdot R_p(f) +
\sum_{\ell\in\text{L}^\text{sdn}} src(\ell ,n) \cdot R_\ell(f) =1
\]
We require the same for the end of the routing path, i.e., routing has to \textbf{end at the destination} of the flow:
\[
\forall f\in\tilde{\text{F}}, \,\,\, \forall n\in\text{N} \,\, \text{ with } \,\, dst(f,n)=1:
\]\[
\sum_{p\in\text{P}} dst(p,n) \cdot R_p(f) +
\sum_{\ell\in\text{L}^\text{sdn}} dst(\ell ,n) \cdot R_\ell(f) =1
\]
Common routing models require only a single routing continuity constraint, but like shown in the flowchart of Figure~\ref{routing}, our model allows two different entities for routing (shown as the intermediate states \emph{OSPF path} and \emph{SDN link}). Thus our model requires \textbf{two different routing continuity constraints} for intermediate nodes $n$. First, we demand that the usage of an OSPF path for the routing of a certain flow always requires the subsequent usage of an SDN link (or the node at the end of that OSPF path is already the destination of the said flow):
\begin{eqnarray*}
\forall f\in\tilde{\text{F}}, \,\,\, \forall p\in\text{P}, \,\,\, \forall n\in\text{N}\text{ with } dst(p,n) = 1:\\
R_p(f) - dst(f,n)
=  \sum_{\ell\in\text{L}^\text{sdn}} R_\ell(f) \cdot src(\ell ,n)
\end{eqnarray*}
Likewise, we demand that the usage of an SDN link for the routing of a certain flow always requires the subsequent usage of another SDN link or alternatively an OSPF path, unless the node at the end of the initial SDN link is already the destination of the said flow:
\begin{eqnarray*}
\forall f\in\tilde{\text{F}}, \,\,\, \forall \ell\in\text{L}^\text{sdn}, \,\,\, \forall n\in\text{N}\text{ with } dst(\ell,n) = 1:\\
R_\ell(f) - dst(f,n)= \,\,\,\,\,\,\,\,\,\,\,\,\,\,\,\,\,\,\,\,\,\,\,\,\,\,\,\,\,\,\,\,\,\,\,\, \\
\sum_{m\in\text{L}^\text{sdn}} R_m(f) \cdot src(m ,n) + \sum_{p\in\text{P}} R_p(f) \cdot src(p ,n)
\end{eqnarray*}

\begin{figure} \begin{center}
\includegraphics[width=6cm]{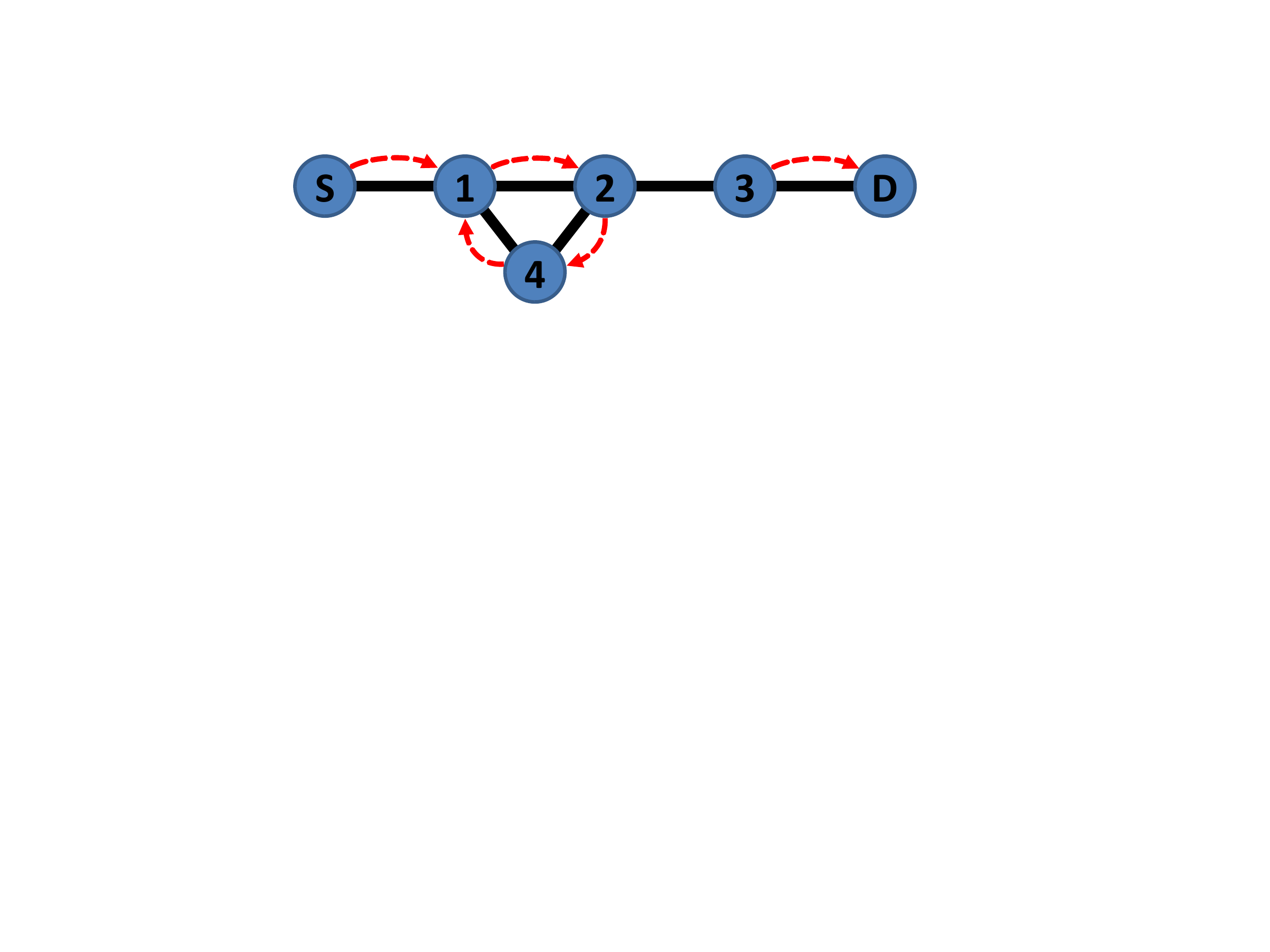}
\caption{Routing loop} \label{routing_loop}
\end{center} \end{figure}
The routing continuity constraints so far may appear to provide a valid routing, however, they do not prevent routing loops like shown in Figure~\ref{routing_loop}. (Please note that it is not distinguished between the two entities \emph{OSPF path} and \emph{SDN link} in Figure~\ref{routing_loop}, as the described problem is of general nature when routing continuity is modeled in ILPs.) While the only valid path from $S$ to $D$ is obviously $S\!\!\rightarrow \!\!1\!\!\rightarrow \!\!2\!\!\rightarrow \!\!3\!\!\rightarrow \!\!D$, the inconsistent routing shown as red dotted arrows is conform to all routing continuity constraints so far: Routing starts at the source ($S\!\!\rightarrow$), at every intermediate node where routing leads to, there is a subsequent node unless we reach the destination ($\rightarrow \!\!1\!\!\rightarrow$, $\rightarrow \!\!2\!\!\rightarrow$ and $\rightarrow \!\!4\!\!\rightarrow$), and the destination is reached ($\rightarrow \!\!D$). The solution is to demand that for each flow, \textbf{any intermediate node can be traversed at most once}, which we model with the following constraint:
\[
\forall f\in\tilde{\text{F}}, \,\, \forall n\in\text{N} \,\, \text{with} \,\, src(f,n)=dst(f,n)=0:
\]\[
\sum_{\ell\in\text{L}^\text{sdn}} R_\ell (f) \cdot \Bigl(src(\ell,n)+dst(\ell,n)\Bigr) \, + \,\,\,\,\,\,
\]\[
\sum_{p\in\text{P}} R_p (f) \cdot \Bigl(src(p,n)+dst(p,n)\Bigr) \leq 2
\]
This assures that the intermediate node $n$, if used for flow $f$, has not more than two routing entities of $f$ in sum that enter and exit the node. Additionally, we need to demand that the \textbf{source and destination nodes never appear as intermediate nodes} along the routing path to avoid loops containing source or destination:
\[
\forall f\in\tilde{\text{F}}, \,\, \forall s,d\in\text{N} \,\, \text{with} \,\, src(f,s)=dst(f,d)=1:
\]\[
\sum_{\ell\in\text{L}^\text{sdn}} R_\ell (f) \cdot \Bigl( dst(\ell,s)+src(\ell,d)\Bigr) \, + \,\,\,\,\,\,
\]\[
\sum_{p\in\text{P}} R_p (f) \cdot \Bigl( dst(p,s)+src(p,d)\Bigr) =0
\]
Finally, we need to constrain that routing through SDN border nodes has to be consistent with the advertised link metrics. This means that for a given sub-domain $\alpha$ and for a given sub-domain-external destination node $d$, we require the \textbf{usage of a set of link metrics} (to be advertised by the SDN nodes in sub-domain $\alpha$) which is consistent with all used OSPF paths from any sub-domain-internal source node to that destination $d$:
\[
\forall p\in\text{P}, \,\,\,\, \forall f\in\tilde{\text{F}}, \,\, \forall d\in\text{N}\text{ with } dst(f,d)=1:
\]
\[
R_p(f) \leq \sum_k LSA_k(d) \cdot \ell sa(k,p)
\]
With this constraint we assure that OSPF path $p$ is used to route flow $f$ only if the usage is consistent with the advertised set of LSAs for the destination of $f$. Finally, as we demand that only a \textbf{single set of LSAs per sub-domain} is advertised per sub-domain-external destination, we also constrain that
\[
\forall d\in \text{N}, \,\,\,\, \forall \alpha \in \text{SD}: \,\,\,\,
\sum_k LSA_k(d) \cdot bel(k,\alpha) = 1
\]
The $lsa$ and $bel$ parameters used in the last two constraints are entirely pre-computed, and determining all valid combinations of OSPF paths for which there has to exist a distinct set of link metrics is computationally complex. In our first ILP model, we had the link metrics advertised by SDN nodes for every single destination as variables, which is straightforward. This, however, turned out to be computationally too expensive even for small network sizes. For space reasons, we can not explain the algorithm that generates the used $lsa$ parameter, but assume here that it is computed efficiently and is available in the model.

\vspace{3mm}\noindent\textbf{Load balancing as objective function:} 
The purpose of the used Traffic Engineering scheme is to load balance the network, that is to increase the \emph{headroom} of the links so that a sudden surge in traffic and large flows are unlikely to congest the network. Therefore, we first define the utilization of a link by
\[
\forall \ell \in \text{L}: \,\,\,\, U_\ell = 
u_\ell + \sum_{f\in\tilde{\text{F}}} \frac{\lambda_f}{C_\ell}\biggl( 
R_\ell(f) + \sum_{p\in\text{P}} R_p(f) \cdot tr(p,\ell) \biggr)
\]
which adds the utilization of each flow (i.e., $f$'s demand divided by $\ell$'s capacity) that is routed by our load balancer via $\ell$ to the utilization caused by all inter-sub-domain (i.e., plain OSPF) flows.

\begin{figure}[t] \center
\includegraphics[width=7.7cm]{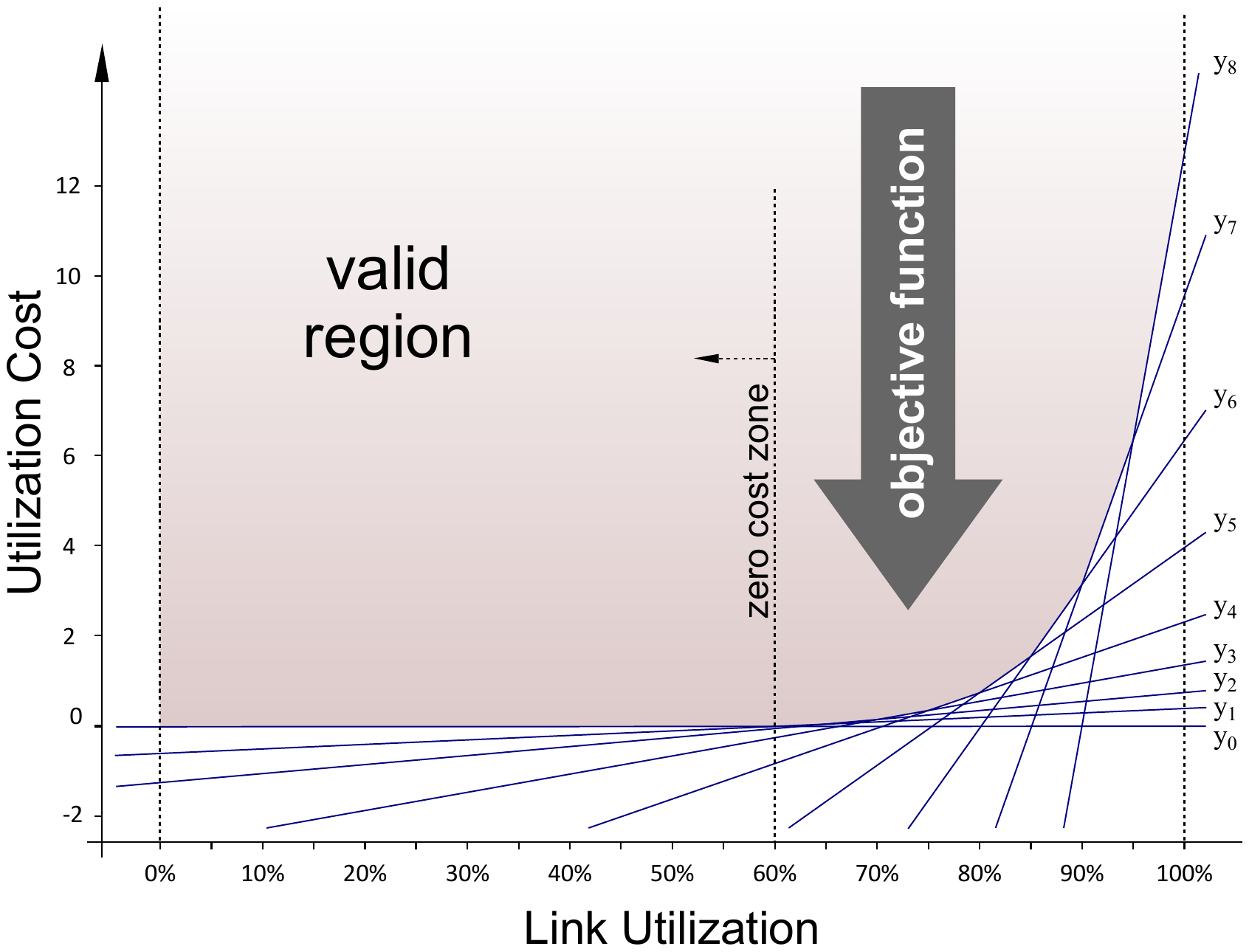}
\caption{Emulation of an exponential cost increase with a piecewise linear function}
\label{costfunction} \end{figure}

We now need to define a load balancing objective and there exist various approaches in the literature: The minimization of the maximum link utilization is one of the common models. However, it was discussed in~\cite{trafficengineering} that this approach has issues with unavoidable bottlenecks: In case of a heavy loaded link that can not be relieved during the optimization, the objective to minimize the maximum link utilization doesn't load balance less loaded links at all. This can be avoided with a cost (i.e., penalty) function that increases exponentially with the link utilization. Every link can then be associated with a cost according to its utilization and the objective of the optimization would be to minimize the total cost in the entire network. As ILP models require linear constraints, an exponentially increasing cost function must be emulated with a piecewise linear function (i.e., with a concatenation of straight lines), like proposed in~\cite{trafficengineering}. The set Y of functions $\{y_0,y_1,...\}$ that we used in this paper is shown in Figure~\ref{costfunction}, and the intention behind it is as follows. Our ISP prefers to route flows over links with utilization below 60\%, and therefore, links with utilization $\leq\!\! 60\%$ generate zero cost. The closer the link utilization gets to 100\%, the more sensitive the link gets to traffic bursts that could cause congestion. The straight lines of Y are fixed such that each of them constitutes the lower bound of the valid cost region for a 5\% section on the x-axis, and the gradient doubles stepwise. The link utilization is evidently bounded by 0\% and 100\% and we constrain the utilization cost for each link to be \emph{greater equal} than all cost functions. The resulting valid solution space is depicted in Figure~\ref{costfunction}. The set Y of linear cost functions was therefore defined as
\[ y_i(U_\ell) = a \cdot U_\ell - b \]
with values $a$ and $b$ given in Table~\ref{lin_param}.
\begin{table}[t]\begin{center}\scriptsize
\begin{tabular}{ c c c c c c c c c c }
\toprule
\textbf{i}	& 0		& 1		& 2		& 3		& 4		& 5		& 6		& 7		& 8 \\
\midrule
\textbf{a}	& 0		& 1		& 2		& 4		& 8		& 16	& 32	& 64	& 128 \\
\textbf{b}	& 0 	& 0.6	& 1.25	& 2.65	& 5.65	& 12.05	& 25.65	& 54.45	& 115.25 \\
\bottomrule
\end{tabular}
\caption{Parameters of the used linear cost functions}\label{lin_param}
\end{center}\end{table}
We accordingly constrain the link utilization cost with:
\[
\forall \ell \in \text{L}, \forall y_i \in \text{Y}: \,\,\,\,
Cost_\ell \geq y_i(U_\ell)
\]
Finally, the objective function is simply:
\[
\text{Minimize}\,\,\,\,\,\, \sum_{\ell\in\text{L}} Cost_\ell
\]

\section{Performance Evaluation}\label{performance-section}
\begin{figure*}[t] \center
\includegraphics[width=\textwidth]{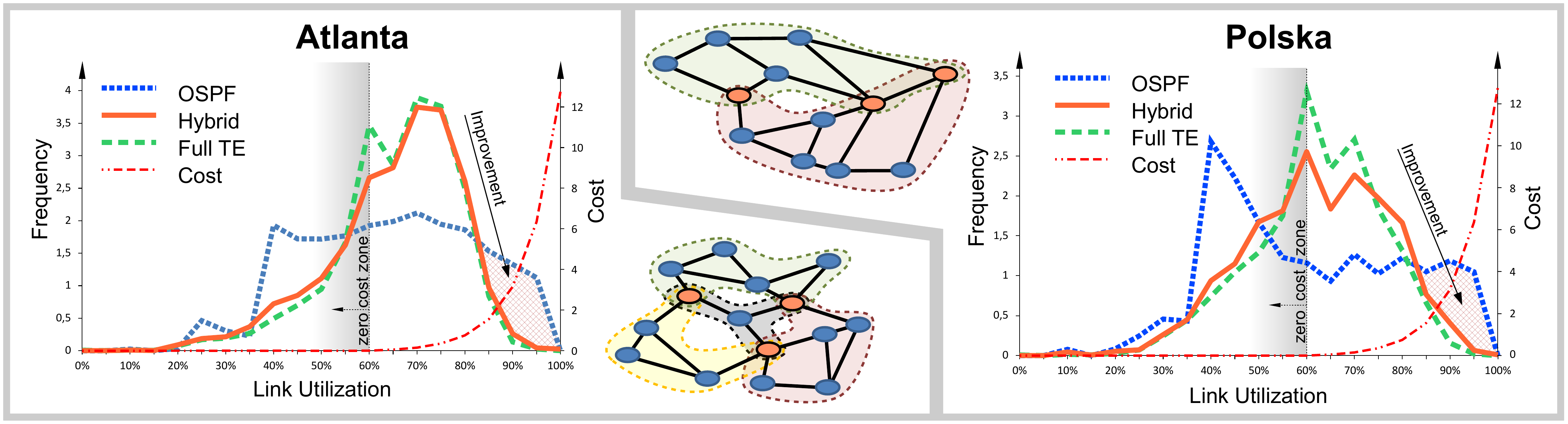}
\caption{Results}
\label{results} \end{figure*}

For our performance analysis, we used the Atlanta and Polska topologies from the SNDlib library~\cite{sndlib}. Figure~\ref{results} shows our resulting graphs, which are the histograms of link utilization, defined as the frequency of the occurrence of a particular link utilization value. Three SDN nodes have been placed according to the heuristic that was explained in Section~\ref{arch-section}, i.e., we chose the nodes providing the maximum number of inter-sub-domain flows. This resulted in a partitioning (also depicted in Figure~\ref{results}) of the Atlanta network into four sub-domains, whereas three SDN nodes in the Polska network partition the topology into two sub-domains.

\par The experiments in both topologies were conducted as follows: For each individual experiment (the results shown are averaged over hundred trials) we initially used the unpartitioned network and assigned uniform link metrics and determined the OSPF least cost paths, which resulted in minimum hop count routing. We then assigned uniformly distributed demands in the range of 1~Gbit/s to 7~Gbit/s to the traffic flows and determined the link loads resulting from the traffic demand and the routing. After that we assigned link capacities greater equal the link loads and assumed that links are available in the following granularity: 10~Gbit/s, 40~Gbit/s, 100~Gbit/s, and 400~Gbit/s. The resulting utilization of links clearly covers a wide range and was used for the \emph{OSPF} plot in both histograms in Figure~\ref{results}, which serves as a kind of "worst case" distribution, as no load balancing is performed.

\par To generate the results for our hybrid SDN/OSPF load balancing scheme, we partitioned the network into sub-domains and deleted all OSPF paths from the parameter set P that are not consistent with our definition in Section~\ref{math-section}, i.e., the paths across sub-domain borders. That scenario was used in each experiment run as input parameters along with the mathematical model explained in Section~\ref{math-section}, which then was optimzed on an Intel Core i7-3930K CPU (6~x~3.2~GHz) using the GUROBI solver. A single experiment run took on average less than three minutes. The results of the load balancing scheme is shown as the \emph{Hybrid} plot in both histograms of Figure~\ref{results}. The utilization cost function is superimposed in both histograms (the \emph{Cost} plot) to highlight how smoothly the load distribution of our scheme mirrors the cost function in the range above 80\% link utilization. 

\par The improvement through load balancing (i.e., the re-routing of flows from highly utilized links) is marked as shaded area and demonstrates the performance of the proposed TE mechanism. As we can see, the link utilizations of 85\% and above occur rarely, while a couple of such heavily loaded links are common in the plain (non-load-balanced) OSPF case. We can also see the occurrence of links utilized above 95\% in the OSPF scheme, whereas our proposed load balancing scheme can prevent congestion in the majority of cases. To show the upper bound on traffic engineering (i.e., the "best case" link utilization distribution), we also computed the optimal routing without any OSPF constraints in each experiment run and added this as the \emph{Full~TE} plot in both histograms. This graph shows the distribution when full TE capabilities are deployed in the network, as if  MPLS or OpenFlow was available at every node. It can be seen that the advantage is marginal, which indicates that even a small number of SDN-enabled routers can enable almost full TE capability. The savings in accumulated link utilization cost (according to the cost function defined in Section~\ref{math-section}) in relation to un-balanced traffic in the plain OSPF case is shown in Table~\ref{costrelation}. Here we can observe some topological differences. The results of our scheme are much closer to the optimum in the Atlanta network than in the Polska network. This reflects the fact that Atlanta network was partitioned into four (instead of only two) sub-domains, which results in a lot more traffic flows that have to cross sub-domain borders, and accordingly more options to traffic engineer the highly utilized links.

\begin{table}[htb]\begin{center}
\begin{tabular}{ c c c }
\toprule
\textbf{Topology} & \textbf{Hybrid SDN/OSPF} & \textbf{Full TE} \\
\midrule
Atlanta & 62.3\% & 67.9\%\\
Polska & 63.9\% & 74.9\%\\
\bottomrule
\end{tabular}
\caption{Saved link utilization cost compared to the unbalanced OSPF case}\label{costrelation}
\end{center}\end{table}

\section{Conclusions}\label{conclusion-section}
In this paper, we proposed a new architecture for a hybrid SDN/OSPF operation, in which a few SDN nodes are used to partition the OSPF domain into multiple sub-domains. We have explained the requirements and assumptions regarding the corresponding network architecture, which suggested that our proposed method can be implemented with ease into common OSPF networks. We showed how OSPF Link State Advertisements can be tuned to optimize the routing. Finally, we illustrated the idea with analysis and numerical results and showed that very few SDN routers are required to achieve results close to the optimum. The low number of required SDN routers also allows that a relatively high number of nodes can remain in the \emph{configure-once-never-touch-again} operation, which is a known and desired feature from OSPF. The results also suggest that the performance depends on the number of sub-domains in which the original topology can be partitioned into. The proposed hybrid management framework can be easily integrated into the ongoing architectural frameworks, such as IETF ABNO and OpenDaylight. The idea is promising and future work will extend the analysis to a wider range of topologies. We are also interested in more suitable and effective partitioning algorithms, especially in larger topologies.  \vspace{5mm}

\textbf{Acknowledgment.} This work has been supported by the German Federal
Ministry of Education and Research (BMBF)
under code 01BP12300A; EUREKA-Project SASER.

\end{document}